%% file: main.tex
\documentclass[acmsmall]{acmart}

\AtBeginDocument{%
  }

\usepackage[]{collab}
\usepackage{enumitem}
\usepackage{balance}
\usepackage{xcolor}
\usepackage{subfigure}
\usepackage{multirow}
\usepackage{multicol}
\usepackage[most]{tcolorbox}
\usepackage{tabularx}
\usepackage{siunitx}
\usepackage{float}
\usepackage{color}
\usepackage{hyperref}
\usepackage{diagbox}
\usepackage{caption}
\usepackage{makecell}

\definecolor{ballblue}{rgb}{0.13, 0.67, 0.8}

\setcopyright{acmlicensed}
\acmDOI{10.1145/3643733}
\acmYear{2024}
\copyrightyear{2024}
\acmSubmissionID{fse24main-p60-p}
\acmJournal{PACMSE}
\acmVolume{1}
\acmNumber{FSE}
\acmArticle{7}
\acmMonth{7}
\received{2023-09-29}
\received[accepted]{2024-01-23}



\newcommand\nm{{LILAC}\xspace}
\newcommand\ourtree{{parsing cache}\xspace}
\newcommand\llmparser{{ICL-enhanced parser}\xspace}

\newcommand{\ie}{{\em i.e.},\xspace}
\newcommand{\eg}{{\em e.g.},\xspace}
\newcommand{\fixedwidth}[1]{{\ttfamily \small #1}}

\collabAuthor{yc}{gray}{Yichen Li}
\collabAuthor{jy}{ballblue}{Jinyang Liu}
\collabAuthor{zb}{teal}{Zhuangbin}
\collabAuthor{jz}{magenta}{Jiazhen Gu}
\collabAuthor{yt}{blue}{Yintong Huo}
\collabAuthor{zh}{green}{Zhihan Jiang}
\collabAuthor{jj}{purple}{Junjie}

\begin{document}


\title{LILAC: Log Parsing using LLMs with Adaptive Parsing Cache}

\author{Zhihan Jiang}
\orcid{0009-0003-1988-6219}
\affiliation{%
  \institution{The Chinese University of Hong Kong}
  \city{Hong Kong}
  \country{China}
}
\email{zhjiang22@cse.cuhk.edu.hk}

\author{Jinyang Liu}
\orcid{0000-0003-0037-1912}
\affiliation{%
  \institution{The Chinese University of Hong Kong}
  \city{Hong Kong}
  \country{China}
}
\email{jyliu@cse.cuhk.edu.hk}

\author{Zhuangbin Chen}
\orcid{0000-0001-5158-6716}
\affiliation{%
  \institution{Sun Yat-sen University}
  \city{Zhuhai}
  \country{China}
}
\email{chenzhb36@mail.sysu.edu.cn}

\author{Yichen Li}
\orcid{0009-0009-8370-644X}
\affiliation{%
  \institution{The Chinese University of Hong Kong}
  \city{Hong Kong}
  \country{China}
}
\email{ycli21@cse.cuhk.edu.hk}

\author{Junjie Huang}
\orcid{0009-0004-6962-5292}
\affiliation{%
  \institution{The Chinese University of Hong Kong}
  \city{Hong Kong}
  \country{China}
}
\email{junjayhuang@outlook.com}

\author{Yintong Huo}
\orcid{0009-0006-8798-5667}
\affiliation{%
  \institution{The Chinese University of Hong Kong}
  \city{Hong Kong}
  \country{China}
}
\email{ythuo@cse.cuhk.edu.hk}

\author{Pinjia He}
\orcid{0000-0003-3377-8129}
\affiliation{%
  \institution{The Chinese University of Hong Kong}
  \city{Shenzhen}
  \country{China}
}
\email{hepinjia@cuhk.edu.cn}

\author{Jiazhen Gu}
\orcid{0000-0002-5831-9474}
\affiliation{%
  \institution{The Chinese University of Hong Kong}
  \city{Hong Kong}
  \country{China}
}
\email{jiazhengu@cuhk.edu.hk}

\author{Michael R. Lyu}
\orcid{0000-0002-3666-5798}
\affiliation{%
  \institution{The Chinese University of Hong Kong}
  \city{Hong Kong}
  \country{China}
}
\email{lyu@cse.cuhk.edu.hk}

\renewcommand{\shortauthors}{Jiang et al.}
\input{Sections/00-abstract}
\titlenote{Jiazhen Gu is the corresponding author.}

\maketitle

\input{Sections/01-introduction}

\input{Sections/02-background}
\input{Sections/03-method}

\input{Sections/04-evaluation}

\input{Sections/05-discussion}
\input{Sections/06-related}
\input{Sections/07-conclusion}




\bibliographystyle{ACM-Reference-Format}
\bibliography{sample-base}

\end{document}

%% file: Sections/00-abstract.tex
\begin{abstract}
Log parsing transforms log messages into structured formats, serving as the prerequisite step for various log analysis tasks.
Although a variety of log parsing approaches have been proposed, their performance on complicated log data remains compromised due to the use of human-crafted rules or learning-based models with limited training data.
The recent emergence of powerful large language models (LLMs) demonstrates their vast pre-trained knowledge related to code and logging, making it promising to apply LLMs for log parsing.
However, their lack of specialized log parsing capabilities currently hinders their parsing accuracy.
Moreover, the inherent inconsistent answers, as well as the substantial overhead, prevent the practical adoption of LLM-based log parsing.

To address these challenges, we propose LILAC, the first practical \underline{L}og pars\underline{I}ng framework using \underline{L}LMs with \underline{A}daptive parsing \underline{C}ache.
To facilitate accurate and robust log parsing, LILAC leverages the in-context learning (ICL) capability of the LLM by performing a hierarchical candidate sampling algorithm and selecting high-quality demonstrations.
Furthermore, LILAC incorporates a novel component, an adaptive parsing cache, to store and refine the templates generated by the LLM.
It helps mitigate LLM's inefficiency issue by enabling rapid retrieval of previously processed log templates.
In this process, LILAC adaptively updates the templates within the parsing cache to ensure the consistency of parsed results.
The extensive evaluation on public large-scale datasets shows that LILAC outperforms state-of-the-art methods by 69.5\% in terms of the average F1 score of template accuracy.
In addition, LILAC reduces the query times to LLMs by several orders of magnitude, achieving a comparable efficiency to the fastest baseline.

\end{abstract}

%% file: Sections/01-introduction.tex
\section{Introduction}

Log messages are generated by logging statements in the source code to record the system events and statuses at runtime.
Modern software systems produce a large volume of log data \cite{yao2021improving,wang2022spine}, facilitating various downstream tasks, such as anomaly detection~\cite{zhang2019robust,zhao2021empirical,zhang2022deeptralog,liu2023scalable,ali2023empirical}, failure troubleshooting~\cite{xu2009largescale,chen2021pathidea} and root cause analysis~\cite{amar2019mining,wang2020root,notaro2023logrule}.
As such, log analysis plays an essential role in the maintenance of software systems.
Log parsing is the first and foremost step in log analysis, which extracts two parts of log messages:
1) \textit{log templates} - constant parts that are explicitly written in logging statements;
2) \textit{log parameters} - dynamic parts that are changeable in different executions.
For example, a logging statement \fixedwidth{``logging.info(f"Starting reading data from \{file\_path\}")''} can generate a sequence of log messages with different \fixedwidth{file\_path}, such as \fixedwidth{``Starting reading data from /etc/data/''}.
In the above example, the log template is \fixedwidth{``Starting reading data from <*>''}, and the log parameter indicates the path of data, \ie \ \fixedwidth{``/etc/data/''}.

Since the source code is generally inaccessible during system maintenance, a wide range of techniques (\ie log parsers)~\cite{vaarandi2003data,nagappan2010abstracting,he2017drain,le2023log} have been proposed to distinguish the templates and parameters from log messages automatically.
Existing log parsers can be categorized into two groups: \textit{syntax-based} and \textit{semantic-based}.
Syntax-based log parsers~\cite{du2016spell,he2017drain,dai2020logram,yu2023brain} utilize specific features or heuristics (\eg log length, word length and frequency) to extract the constant parts of log messages as templates.
In contrast, semantic-based log parsers~\cite{huo2023semparser,liu2022uniparser,le2022log,li2023did} employ deep learning models to learn semantics and system-specific patterns from labeled log data so as to parse new log messages.

Unfortunately, recent benchmark studies~\cite{khan2022guidelines,jiang2023large,petrescu2023log} have revealed that the performance of existing log parsers in practice remains unsatisfactory.
On the one hand, syntax-based log parsers heavily rely on crafted rules, while a significant performance degradation would happen when the log data deviate from the established rules.
On the other hand, the deep learning models adopted by semantic-based log parsers are trained by limited labeled log messages.
When parsing more complicated log messages that have different features from the training data, the models may fail to understand semantics and extract templates.

To address these limitations, we propose to leverage the powerful large language models (LLMs) to achieve effective log parsing.
LLMs are trained by vast amounts of text data related to
code~\cite{yang2023code,peng2023generative} and logging~\cite{mastropaolo2022using,li2023exploring}, thus having the potential to understand log messages comprehensively. 
For example, when processing a log message \texttt{``Process f3e2 write to /etc/smartd.conf failed.''}, the LLM can accurately discern that \texttt{``f3e2''} and \texttt{``/etc/smartd.conf''} are parameters recording the process ID and the file path.
Moreover, this process does not require manually designed rules (\eg regular expressions and delimiters), which makes LLMs promising components for log parsing. 
However, designing a practical LLM-based log parsing approach still faces the following challenges:

(1) Lack of specialized capability. 
LLMs are not specialized in log parsing. 
Although LLMs have a wealth of general knowledge through pre-training, they are not fine-tuned (\eg instruction tuning~\cite{wei2021finetuned} and reinforcement learning with human feedback~\cite{macglashan2017interactive}) for the log parsing task.
Hence, the performance of directly querying LLMs to parse log messages may be compromised~\cite{le2023evaluation,mudgal2023assessment}.

(2) Inconsistent outputs of LLMs.
As revealed by recent studies~\cite{du2023improving,mundlerself,peng2023check,mudgal2023assessment}, LLMs may produce unstable outputs.
In terms of log parsing, LLMs may generate different templates for log messages with the same template.
This inconsistency will lead to a decline in grouping accuracy, a critical factor for certain downstream tasks, such as log compression~\cite{li2023logshrink,rodrigues2021clp} and anomaly detection~\cite{zhang2019robust,le2022log}.

(3) Huge overhead of employing LLMs.
LLMs have billions of weights and require huge computing resources (\eg GPUs) for inference.
Therefore, compared to traditional parsing tools, the overhead of querying LLMs (\eg inference time and network latency) is notably high~\cite{dettmers2022llm,wang2023tabi}.
Considering that modern software systems can produce tens of gigabytes of logs per hour~\cite{zhu2019tools,wang2022spine,li2023logshrink}, directly employing LLMs for log parsing is impractical.

To tackle the aforementioned challenges, we propose \nm, a \textbf{\underline{L}}og pars\textbf{\underline{I}}ng 
framework using \textbf{\underline{L}}LMs with \textbf{\underline{A}}daptive parsing \textbf{\underline{C}}ache.
\nm consists of two main components, the \textit{\llmparser} and the adaptive \textit{\ourtree}.
The \llmparser is designed to accurately parse queried log messages, while the \ourtree stores and adaptively refines the generated templates to ensure both efficiency and consistency.
In particular, the \llmparser leverages the in-context learning (ICL) capability to adapt LLMs to parse diverse log data. It first obtains high-quality demonstrations using the proposed effective and efficient \textit{candidate sampling} and \textit{demonstration selection} algorithms, and then utilizes the designed prompt format to guide the LLM to parse log messages accurately.
The design of the \ourtree targets to address the issues of inconsistent outputs and huge overhead associated with LLMs.
By prioritizing the cache matching operation, \nm can avoid duplicated queries to LLMs, thereby enhancing the parsing efficiency.
Moreover, the cache updating operation can adaptively refine the potential erroneous templates within the \ourtree to mitigate the inconsistency of LLMs.

We have conducted a comprehensive evaluation on public large-scale log datasets of Loghub-2.0~\cite{jiang2023large} from the LogPAI team~\cite{zhu2019tools}.
The results show that \nm achieves the highest average accuracy on all performance metrics, outperforming state-of-the-art baselines by 66.8\% and 69.5\% for the F1 score of grouping and template accuracy, respectively.
Furthermore, \nm exhibits remarkable robustness across diverse log datasets, consistently maintaining high performance when integrated with various language models.
With regards to efficiency, \nm has achieved a speed comparable to the most efficient baseline, Drain~\cite{he2017drain}, significantly reducing the overhead of querying LLMs.

The main contributions of this work are summarized as follows:

\begin{itemize}[leftmargin=*, topsep=0pt]
    \item To the best of our knowledge, we propose the first practical LLM-based log parsing framework named \nm.
    With effective and efficient candidate sampling and demonstration selection algorithms, \nm exploits the ICL capability of LLMs, enabling accurate and robust log parsing.
    \item We introduce an adaptive \ourtree and design cache operations to mitigate the inefficiency and instability issues associated with the application of LLMs for log parsing.
    \item We extensively evaluate \nm on public large-scale datasets.
    The results show that \nm outperforms state-of-the-art methods in terms of accuracy while also achieving high efficiency. 
    \item The source code of \nm is publicly available at \href{https://github.com/logpai/LILAC}{https://github.com/logpai/LILAC} to benefit both practitioners and researchers in the field of log analysis.
\end{itemize}

%% file: Sections/02-background.tex
\section{Background and Motivation}

\vspace{-1pt}
\subsection{Log Parsing}

Log parsing aims to convert semi-structured log messages into structured data, \ie extracting both the constant parts (\ie log templates) and the dynamic parts (\ie log parameters) from log messages.
A straightforward method involves matching raw log messages with corresponding logging statements within the source code~\cite{pecchia2015industry,schipper2019tracing}.
However, this strategy is impractical when the source code is inaccessible, such as commercial software.
Consequently, a variety of data-driven log parsers without requiring access to the source code have been proposed in the literature~\cite{jiang2008abstracting,he2017drain,yu2023self}.
These log parsers can be categorized into two groups: syntax-based ones and semantic-based ones.

Unfortunately, recent studies have underscored that existing log parsers struggle when handling diverse log data~\cite{khan2022guidelines,petrescu2023log,jiang2023large}.
On the one hand, syntax-based log parsers heavily rely on pre-designed features and rules (\eg regular expressions), requiring a substantial amount of domain-specific knowledge.
This limitation leads to a compromised performance when processing log data that does not adhere to these established rules.
For instance, Drain~\cite{he2017drain}, a leading syntax-based log parser, employs heuristics based on the assumption that all log parameters within specific templates possess an identical number of tokens.
This assumption can lead to errors in the parsed templates when the parameter length exhibits flexibility.
On the other hand, semantic-based log parsers typically adopt deep learning models to utilize the semantics within log messages.
Hence, they are inherently limited by the quantity and quality of labeled data available for model training or tuning.
This limitation can lead to a substantial degradation in their performance when processing complex and large-scale log data~\cite{jiang2023large,xu2023prompting}.
Additionally, the log messages generated in production systems are continually evolving, resulting in ever-changing characteristics of log data~\cite{wang2022spine,xu2023prompting}.
This evolution may render training-based or tuning-based methods non-adaptive to the changes in log data, subsequently leading to unsatisfying practical performance.

\vspace{-4pt}
\subsection{Large Language Models}
\label{sec: LLM-ICL}

Large Language Models (LLMs) have demonstrated remarkable performance in the field of natural language processing.
These models generally adopt the Transformer~\cite{vaswani2017attention} architecture and are trained on extensive corpora using self-supervised objectives.
LLMs are characterized by their large sizes, \eg the standard GPT-3 model~\cite{brown2020language} has 175 billion parameters.
Recently, many studies (\eg SPINE~\cite{wang2022spine} and Hue~\cite{xu2023hue}) have introduced the ``human-in-the-loop'' concept, indicating the need for external knowledge for effective log analysis.
Given that LLMs already possess a substantial amount of pre-trained knowledge, it is promising to utilize LLMs for log parsing.

However, how to effectively apply LLMs to downstream tasks has emerged as a vital research topic.
A common approach involves fine-tuning the model and updating the parameters using specific downstream datasets.
Nonetheless, this method demands considerable computational resources and high-quality data, making it less feasible in specific scenarios.
In contrast, \textit{in-context learning} (ICL) presents an innovative alternative to utilize LLMs to perform downstream tasks~\cite{dong2022survey,liu2023pre}.
Specifically, in the ICL paradigm, the prompt to query LLMs typically comprises three parts:
(1) \textit{Instruction}: description of the specific task; 
(2) \textit{Demonstrations}: several examples, \ie pairs of queries and corresponding ground-truth answers;
(3) \textit{Query}: the query that requires an answer from LLMs.
Such a prompt can let LLMs gain task-specific knowledge by learning the input-output relationship of the task.
Recent studies have demonstrated that ICL can aid LLMs in achieving remarkable performance in a variety of tasks such as logic reasoning~\cite{wei2022chain} and fact retrieval~\cite{he2022rethinking}.
Therefore, in this paper, we intend to adopt LLMs with the ICL paradigm to achieve effective log parsing.

\subsection{Challenges of Log Parsing with LLMs} 

Although utilizing LLMs for log parsing presents significant potential and some recent work~\cite{liu2023logprompt,le2023evaluation,xu2023prompting} has investigated the LLM-based log parsing, these studies fall short in addressing the following three critical challenges, which prevent their practical adoption.

\begin{itemize}[leftmargin=*]
    \item \textit{Specialization.}
    Although LLMs are imbued with a large volume of pre-trained knowledge, they are not specialized in the log parsing task.
    As a result, directly querying LLMs to perform log parsing could potentially result in a compromised performance.
    The ICL paradigm can facilitate the adoption of LLMs to log parsing without tuning.
    Specifically, the demonstrations within the prompt can impart task-specific knowledge to LLMs by the correlation between input and output.
    In practice, the initial phase of ICL involves sampling a small set of candidate log messages, from which the demonstrations for each query will be selected.
    Given the huge volumes and imbalanced frequencies~\cite{wang2022spine,khan2022guidelines,jiang2023large} of logs in real-world systems, it is quite challenging to select diverse candidate log messages for effective ICL.
    Though some sampling algorithms exist for few-shot log parsing~\cite{xu2023prompting, le2023log}, all of them require pairwise computation between log messages or adopt random sampling, which can hardly choose diverse candidates efficiently.
    Hence, the efficient sampling of a set of diverse candidate log messages to enable effective ICL still presents a challenge.
    \item \textit{Consistency.}
    Despite the strong capabilities of understanding and generating texts, LLMs may produce unstable answers, which has been identified and discussed in recent studies~\cite{zheng2023does, mundlerself, peng2023check}.
    Furthermore, due to the limitation of LLMs in parsing based solely on the semantics within a single query, they may exhibit inconsistency in determining whether a particular token is a parameter.
    These may lead LLMs to produce templates that are either more broad or more specific when parsing two log messages that share the same template but have distinct parameter values.
    For example, when parsing two distinct log messages, \fixedwidth{``User root failed to kill the process 0xF28A''} and \fixedwidth{``User user1 failed to kill the process 0x6C37''} individually, inconsistency may arise in the answers of the LLM. 
    Specifically, the LLM may identify \fixedwidth{``root''} in the first log message as a constant token while identifying \fixedwidth{``user1''} in the second log message as a parameter.
    These inconsistent templates can precipitate a decrease in grouping accuracy, which would impact downstream tasks such as log compression and anomaly detection.
    Therefore, mitigating the inconsistency of LLMs to generate consistent log templates is yet to be resolved.
    \item \textit{Efficiency.}
    Given that real-world systems generate substantial volumes of log data, \eg tens of gigabytes per hour~\cite{zhu2019tools,wang2022spine,jiang2023large}, log parsers should process high volumes of data efficiently, \eg millions of log messages per minute. 
    Therefore, efficiency is a critical aspect of practical log parsers. 
    Since LLMs have billions of weights and require extensive resources (\eg high-performance GPUs) for inference, they are typically deployed on high-performance servers and provide query interfaces.
    Compared to traditional local-deployed parsing tools,  utilizing LLMs inevitably introduces considerable overhead, including inference time and network latency~\cite{dettmers2022llm,wang2023tabi}.
    Existing work~\cite{le2022log,le2023evaluation,xu2023prompting} employs LLMs or other language models to process each log message individually, which is hard to meet the practical efficiency demands~\cite{jiang2023large,mudgal2023assessment}.
    Consequently, how to achieve efficient LLM-based log parsing remains a challenge to be addressed.
\end{itemize}

%% file: Sections/03-method.tex
\section{method}

\begin{figure*}[]
    \vspace{-5pt}
    \centering
    \includegraphics[width=\textwidth]{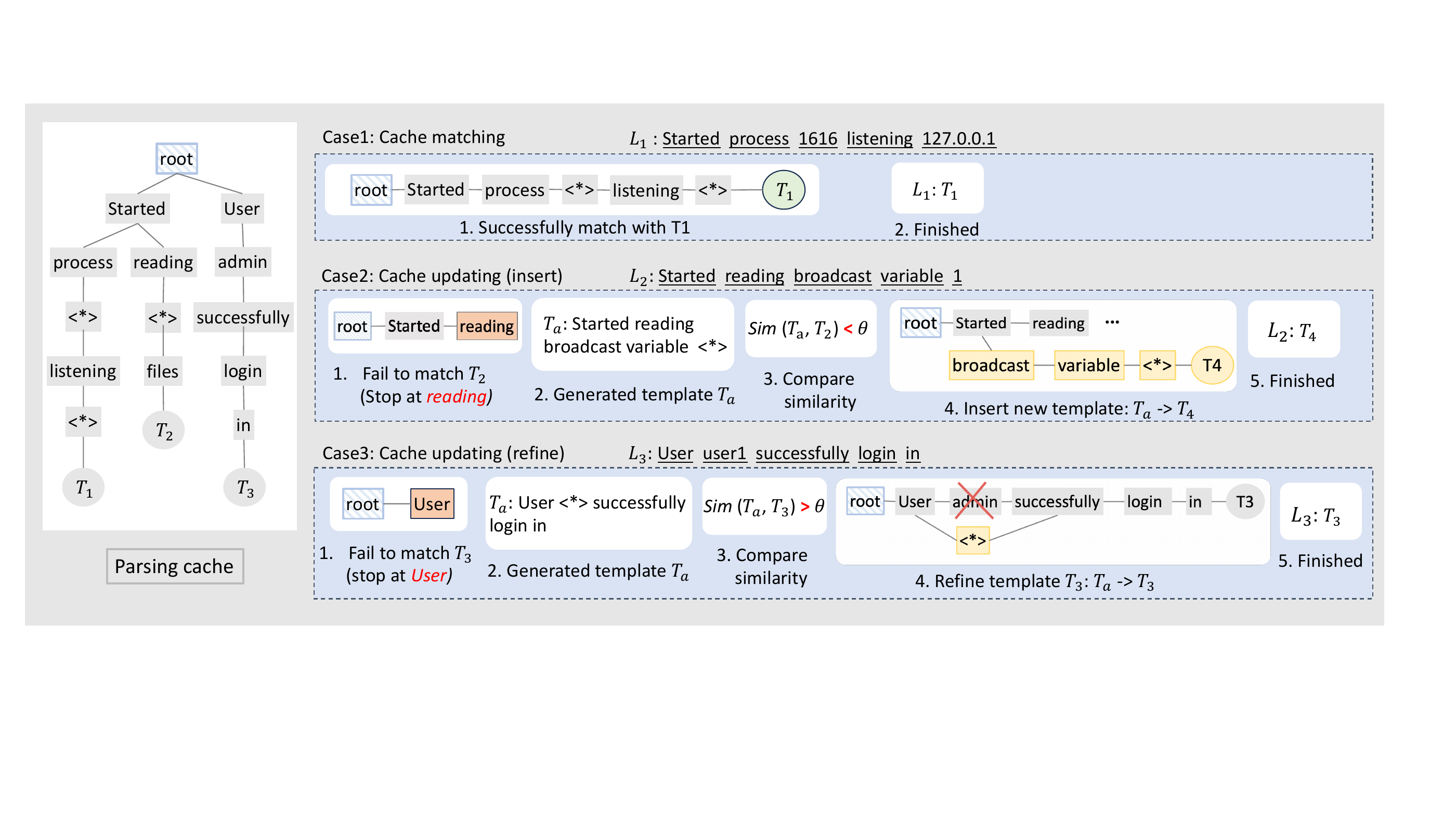}
    \caption{The overall workflow of \nm.} 
    \label{fig: framework}
    \vspace{-10pt}
\end{figure*}

\subsection{Overview}

In this section, we introduce \nm, a log parsing framework using LLMs with adaptive \ourtree.
\nm consists of two components: the \textit{\llmparser} and the adaptive \textit{\ourtree}.
To specialize the LLM in log parsing and adapt it to various log data, the \llmparser utilizes the ICL paradigm.
Specifically, an efficient candidate sampling algorithm is performed to choose a candidate set of diverse log messages, from which effective demonstrations can be selected for the LLM.
To address the inefficiency associated with the utilization of LLMs, \nm introduces a novel component, the \ourtree, to store the parsed templates.
Such a design is motivated by the following observation:
The number of log templates is several orders of magnitude smaller than the number of log messages in real-world systems~\cite{liu2019logzipIC,wang2022spine,jiang2023large}.
For instance, the datasets in Loghub-2.0~\cite{jiang2023large} contain over 50 million log messages, yet the total number of log templates is fewer than 3,500.
Hence, through caching and matching the parsed log templates, \nm can avoid duplicate LLM queries and significantly improve the parsing efficiency.
Moreover, to ensure the consistency of the parsed results, \nm adaptively refines the stored log templates within \ourtree based on the newly generated templates from the \llmparser.

Fig.~\ref{fig: framework} overviews the workflow of \nm.
For each log message to parse, \nm first performs the cache matching operation to check whether its corresponding template is already stored in the \ourtree.
If so, \nm directly used the matched template as the parsed result of this log message, thereby preventing duplicate queries of the LLM.
Otherwise, the cache matching operation will retrieve several relevant templates from the \ourtree, which exhibit a high degree of correlation with the input log message.
Since these relevant templates may be erroneous templates caused by mistakes of the LLM, \nm will record them for the subsequent adaptive updating.
Then, the \llmparser selects high-quality demonstrations from the sampled candidate set to form the designed prompt.
This prompt is then used to query the LLM to extract the template for this log.
Based on both the generated template and its relevant templates, in the cache updating operation, \nm will adaptively determine whether to insert the generated template as a new template to the \ourtree, or to refine an existing relevant template to achieve more precise and consistent parsed results.

\input{Sections/03.5-method-part1}

\input{Sections/03.5-method-part2}

%% file: Sections/03.5-method-part1.tex
\subsection{\llmparser}

\begin{figure*}[htbp]
    \centering
    \vspace{-5pt}
    \includegraphics[width=\textwidth]{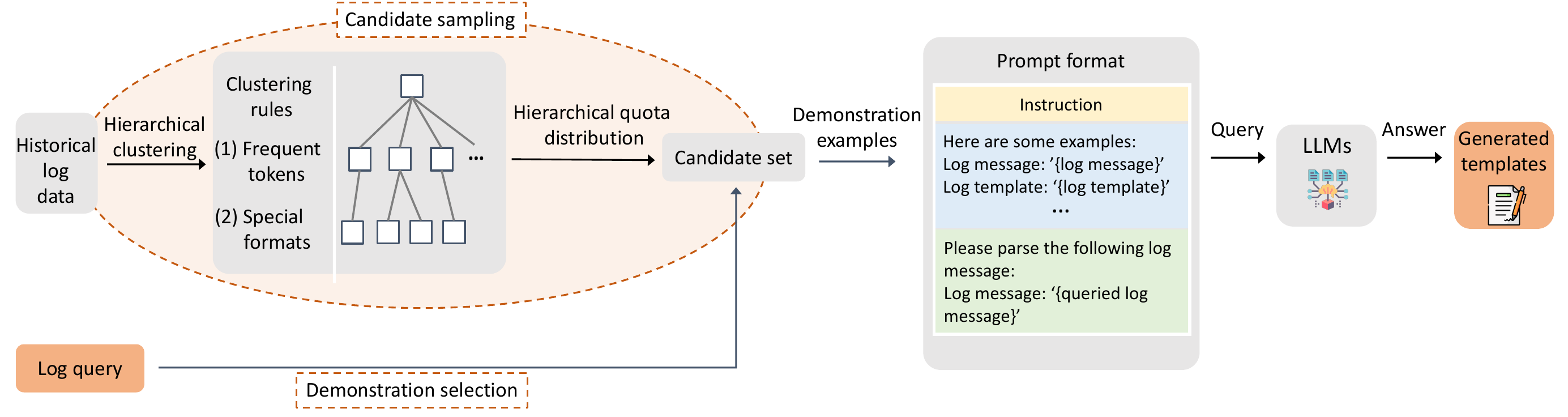}
    \caption{The workflow of \llmparser.} 
    \label{fig: component-2}
    \vspace{-5pt}
\end{figure*}

Fig.~\ref{fig: component-2} presents the overall design of the \llmparser adopted by \nm.
It employs the ICL paradigm to adapt the LLM to log parsing task without resource-intensive model tuning.
Furthermore, it leverages system-specific features within demonstrations to facilitate more accurate log parsing.
In particular, we propose an effective and efficient \textit{candidate sampling} algorithm, along with a \textit{demonstration selection} algorithm, to obtain high-quality examples for effective ICL.
Specifically, \nm first performs the hierarchical candidate sampling algorithm to sample a small set of diverse and representative candidate log messages.
During the online parsing, for each queried log, \nm utilizes the KNN-based demonstration selection algorithm to choose similar demonstration examples.
These demonstrations are integrated into the prompt following the designed format.
Lastly, the \llmparser inputs the prompt to the LLM and obtains the generated templates.

\subsubsection{Candidate Sampling}
\label{sec: candidate sampling}
A typical application of the ICL paradigm involves initially sampling a small set of candidate log messages from produced log data in the system.
It is crucial to ensure that the candidate set is diverse to mitigate the potential risk of inductive bias~\cite{le2023log,xu2023prompting}, since disproportionate demonstrations could cause the LLM to overfit to a specific example.
Furthermore, these candidates ought to be representative, \ie they should be capable of representing more log messages within the log data to provide the LLM with more system-specific characteristics.
In real-world applications, this sampling procedure can initially be performed on the historical log data and be executed periodically to maintain a dynamic candidate set, thereby facilitating adaptation to the continuous evolution of log data.
Therefore, the efficiency of this sampling process is essential.

In the \llmparser, we propose a hierarchical sampling algorithm to extract a small, diverse, and representative set of log messages from substantial log data, as illustrated in the left part of Fig.~\ref{fig: component-2}.
This algorithm consists of two phases, \textit{hierarchical clustering} and \textit{hierarchical quota distribution}.
It first groups the entire log data into hierarchical clusters based on the characteristics of log messages.
Each cluster encompasses log messages with highly similar features, whereas log messages within different clusters exhibit divergent characteristics.
Then, a hierarchical quota distribution approach is performed to select candidates from different clusters while assigning distinct priorities to each cluster based on its number of log messages.

\noindent
\textbf{Hierarchical clustering.}
Inspired by previous research~\cite{nagappan2010abstracting,liu2019logzipIC,jiang2023large}, we first utilize the top-K frequent tokens to group log messages.
The intuition is that log messages that share the same frequent tokens are more likely to have the same templates.
Specifically, we first tokenize each log message and then calculate all token frequencies.
During the above process, stop words in the Scipy library~\cite{scipy} are excluded to eliminate irrelevant tokens.
For each log message, tokens with top-K frequencies are selected, which form the basis for their categorization into different \textit{coarse-grained clusters}.
In other words, all log messages within the same coarse-grained clusters share the same top-K frequent tokens.

However, solely utilizing frequent tokens is insufficient to differentiate log messages with varying characteristics, \ie log messages that share the same top-K frequent tokens may correspond to different log templates.
Thus, we leverage the special characters (\ie characters that are not alphabets, numerals, or white space) to reflect the features of log messages, defining the set of special characters in a log message as its \textit{special format}.
Log messages originating from the same template typically share an identical special format.
This is because the special characters in the constant parts (\ie the template) of a log message are invariably identical, and those in the dynamic parts (\ie the parameter) are generally congruent.
For instance, the special format of \fixedwidth{``Received block: blk\_358 of size 6710 from /127.0.0.1''} is \fixedwidth{\{`:',`\_',`.',`/'\}}.
For other log messages that share the same template, such as \fixedwidth{``Received block: blk\_729 of size 8199 from /127.0.0.2''}, they would have an identical special format.
Therefore, we use the special formats of log messages to perform fine-grained clustering.
In detail, log messages in each coarse-grained cluster are further divided according to their special formats and constitute \textit{fine-grained clusters}, wherein all log messages not only have identical top-K frequent tokens but also share the same log format.

\noindent
\textbf{Hierarchical quota distribution.}
In this phase, we aim to choose diverse and representative log messages as candidates from the fine-grained clusters.
The core idea is to hierarchically distribute the quota of $K_s$ candidates across all fine-grained clusters as evenly as possible to enhance diversity.
Moreover, we assume that clusters with a larger number of log messages are more representative.
Hence, in situations where achieving an equitable distribution is unattainable, priority is given to fine-grained clusters with more log messages.

Initially, we distribute the quota of $K_s$ candidates uniformly across all coarse-grained clusters.
Subsequently, within each coarse-grained cluster, we arrange all fine-grained clusters in descending order based on their priorities determined by the number of log messages they contain.
Given a coarse-grained cluster that has been allocated $K_c$ quotas and contains $n$ sorted fine-grained clusters, denoted as \{$f_1$, $f_2$, $\cdots$, $f_{n}$\}, the quota assigned to cluster $f_i$ is as follows:
\begin{equation*}
S(f_i) =
\begin{cases}
\lfloor \frac{K_c}{n} \rfloor + 1 &\quad\text{if } i \le (K_c \bmod n)\\
\lfloor \frac{K_c}{n} \rfloor &\quad\text{otherwise}\\
\end{cases}
\end{equation*}

Recall that the design of the \llmparser in \nm is intended to leverage the ICL paradigm in few-shot scenarios, which suggests that the number of sampled candidates, $K_s$, is typically small.
This means, in most cases, the number of fine-grained clusters surpasses the number of sampled candidates, \ie $N_f > K_s$.
Hence, the quota allocation for each fine-grained cluster is also typically small (\eg 0 to 2).
Lastly, we randomly select the assigned number of candidate log messages within each fine-grained cluster to ensure high efficiency.

\subsubsection{Demonstration Selection}
\label{sec: demonstration selection}

During the parsing process, to mitigate the interference of irrelevant information and enhance log parsing accuracy, we need to further select $k$ demonstration examples from the $K_s$ candidates to construct the prompt for ICL.
These demonstration examples should exhibit similarity to the queried log message, providing the LLM similar patterns and semantics within the examples to parse this log message accurately~\cite{xu2023prompting,gao2023constructing}.

\nm adopts k-Nearest Neighbors (kNN), a simple yet effective algorithm to select demonstration examples.
For each queried log message $l$, we compute the similarities between it and all candidate log messages, \ie $sim(l, s_i)$, $i \in [1, K_s]$.
We propose to measure the similarity between two log messages based on both tokens and special formats.
In specific, given a log message $l$, we extract the characters of the tokens that are derived from $l$ and the special characters within $l$, to form the \textit{feature set} of $l$, \ie $F(l)$.
Based on this, we can calculate the value of $sim(\cdot)$ of two log messages by using the Jaccard similarity~\cite{jaccard} of their feature sets, \ie $sim(l_1, l_2) = \frac{|F(l_1) \ \cap \ F(l_2)|}{|F(l_1) \ \cup \ F(l_2)|}$.
After computing all similarities, we select log messages from the candidate sets that exhibit the top-$k$ highest similarities.
These log messages, characterized by similar tokens and special characters to the queried log, are instrumental in aiding the LLM to comprehend the semantics and formats embedded within them.

\subsubsection{Query Design}

\begin{figure*}[htbp]
    \vspace{-5pt}
    \centering
    \includegraphics[width=\textwidth]{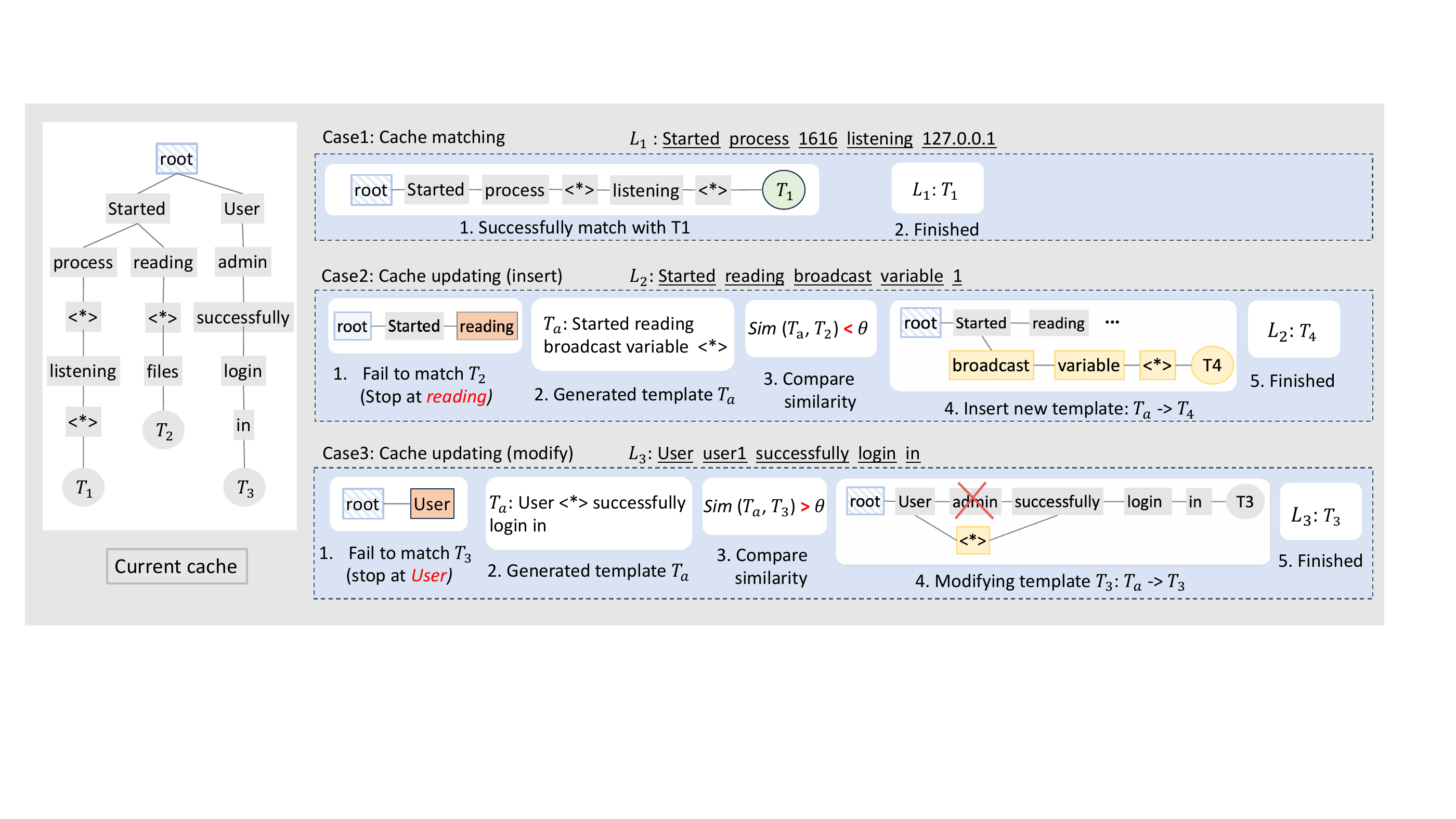}
    \caption{The demonstration of our prompt design.} 
    \label{fig: prompt}
    \vspace{-5pt}
\end{figure*}

Following previous work~\cite{xu2023prompting,le2023evaluation}, we design and use the prompt format, as depicted in Fig.~\ref{fig: prompt}, to query the LLM to generate the log template for an individual log message.
Specifically, the prompt encompasses the following three parts.

\begin{enumerate}[leftmargin=*, topsep=0pt]
    \item \textit{Instruction.}
    To provide the LLM with more task-specific information, we employ an instruction that briefly introduces the task, the concept of log parsing, and the formats of input and output.
    \item \textit{Demonstration Examples.}
    Subsequently, we integrate several demonstrations, chosen by the demonstration selection algorithm, into the prompt.
    Each demonstration includes a pair of one log message and its log template.
    Since recent work~\cite{zhao2021calibrate,gao2023constructing} has pinpointed that LLMs with ICL are more prone to be influenced by the examples that are closer to the query, we arrange the demonstration examples in \textit{ascending order} of similarity to the queried log, \ie those of higher similarities closer to the queried log.
    This is based on the intuition that examples with higher similarity may encompass more information pertinent to parsing the queried log.
    \item \textit{Queried Log.}
    Last, we present the content of the log message to query the LLM.
\end{enumerate}

Guided by the instruction and demonstration examples, the LLM could more precisely answer the log template of the queried log in the prompt, adhering to the correct format.

%% file: Sections/03.5-method-part2.tex
\subsection{Adaptive Parsing Cache}

\begin{figure*}[htbp]
    \vspace{-5pt}
    \centering
    \includegraphics[width=\textwidth]{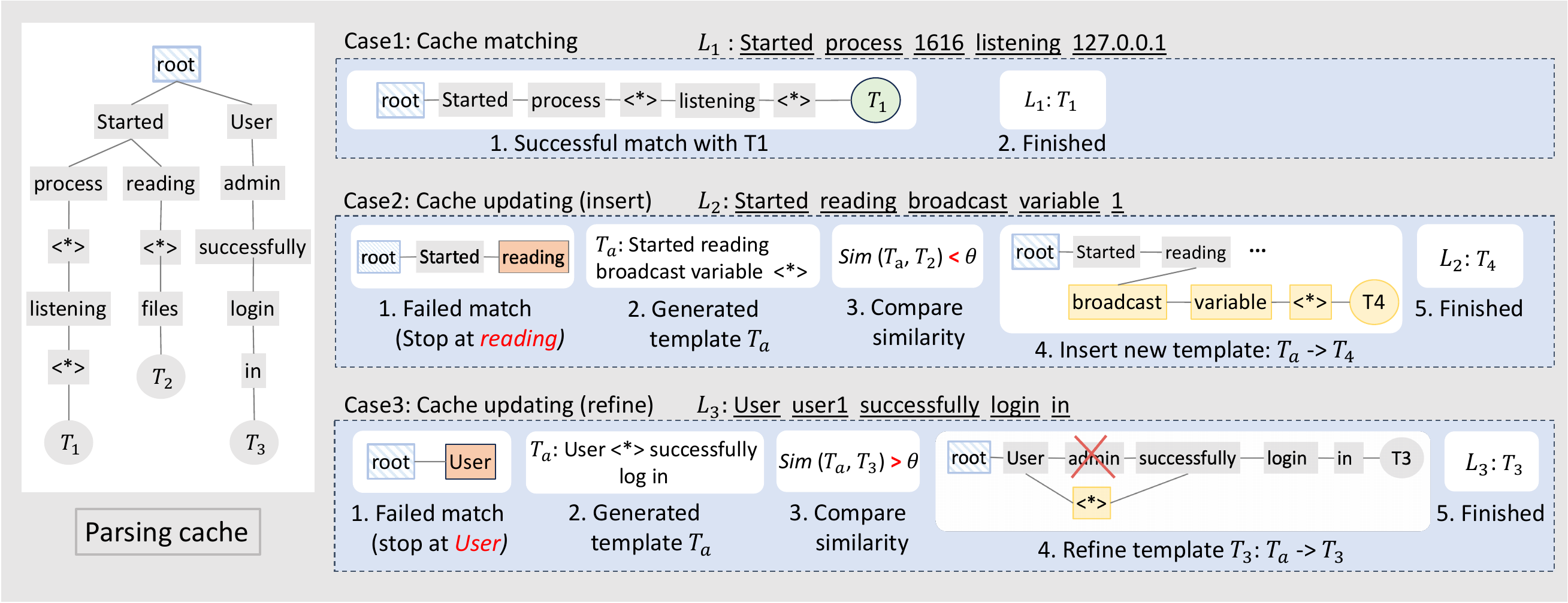}
    \caption{The demonstration cases of cache matching and updating operations for the \ourtree.}
    \label{fig: component-1}
    \vspace{-5pt}
\end{figure*}

The adaptive \ourtree is designed to guarantee the efficiency and consistency of \nm.
Specifically, \nm adopts a tree structure to store the generated templates of the LLM, serving as the \ourtree.
The left part of Fig.~\ref{fig: component-1} demonstrates an example of \ourtree, which stores three log templates.
In the \ourtree, all generated log templates are tokenized into a list of tokens, which are stored in the tree from top to bottom. 
Each intermediate node in the tree represents a token, with the \fixedwidth{``<*>''} denoting the wildcard token that can match any length of tokens.
Each leaf node of the tree represents a unique log template, which corresponds to the string obtained by concatenating all tokens contained in all intermediate nodes on the unique path from the root node to the leaf node.
This tree structure allows for efficient storage and parallel retrieval of log templates.
To retrieve a specific template, only one single traversal from the root to the leaf node is required, without the necessity to check each template sequentially (Sec.~\ref{sec: cache_matching}). 
Moreover, the tree structure can directly reflect the similarity among log templates, \ie templates within the same subtree share a common prefix. 
This can aid in filtering relevant templates of specific log messages, which will be further used for cache updating operation (Sec.~\ref{sec: cache_updating}).

Based on the tree structure of \ourtree, we further design two cache operations, \ie \textit{cache matching} and \textit{cache updating}.
The cache matching operation is used to determine whether the template of the input log message has already been stored in \ourtree.
The cache updating operation is designed to adaptively update templates stored in the \ourtree when the cache matching fails and a new template is generated by the \llmparser.
Next, we illustrate the design details of these cache operations.

\subsubsection{Cache Matching}
\label{sec: cache_matching}

Given a new log message, \nm first checks whether the corresponding template has been stored in the \ourtree through the cache matching operation, which can reduce duplicate queries to LLM and improve parsing efficiency.
To match the input log message with the \ourtree, we first split the log content into a series of tokens by delimiters.
Then, these tokens are read sequentially from the first to the last, with each being compared to intermediate nodes within the tree structure of \ourtree.
Specifically, for the initial token, a search is conducted to verify its presence in the second layer of the tree since the first layer is the empty root node.
If a match is found for the first token, the process continues with the second token and the children of the matched node.
This procedure persists until all tokens have been read or until no further tokens can be matched.
It is worth noting that the wildcard token \fixedwidth{``<*>''} represents parameters of variable length, thus it can match more than one token.
Consistent with existing work~\cite{liu2019logzipIC}, we employ recursive processing to match the wildcard.
Furthermore, to prevent overly broad matching, a limit is also imposed on the maximum number of tokens that a single \fixedwidth{``<*>''} can match.

After the matching process, reaching a leaf node indicates an exact match of the template represented by the leaf node and the input log message, \ie the template of this log is stored in the \ourtree.
Hence, there is no necessity to query the \llmparser again, and the id of the leaf node is recorded as the template id of this log message.
As shown in Case1 of Fig.~\ref{fig: component-1}, when the log message $L_1$ successfully matches the template $T_1$ stored in the \ourtree, \nm directly marks $T_1$ as the parsed template of $L_1$.
In some special cases, multiple matched templates may be returned.
Consistent with previous research~\cite{he2017drain,jiang2023large}, the template with the longest constant parts is selected, as it can match more non \fixedwidth{``<*>''} characters, thereby indicating a higher likelihood of it being the template for this log message.

If no leaf node is reached after the recursive matching process, it will terminate at one or more internal nodes, referred to as \textit{stop nodes}.
In such cases, all templates within the subtrees of stop nodes form a list of relevant templates, denoted as $[T_1, T_2, \cdots, T_n]$. 
For example, in Case2 of Fig.~\ref{fig: component-1}, the cache matching process stops at the \fixedwidth{``reading''} node, so the only template $T_2$ within the subtree is the relevant template.
These relevant templates share a portion of prefixes identical to the current log message without an exact match.
The failed matching may be caused by:
(1) It is the first parsed log message of its respective template.
(2) The LLM produces erroneous templates for previous log messages with the same template.
To discriminate the above two circumstances, the cache matching operation will return relevant templates for subsequent cache updating operation.

\subsubsection{Cache Updating}
\label{sec: cache_updating}

When the cache matching of a specific log message fails, \nm will query the \llmparser to generate the template $T_a$. 
Although the LLM can correctly parse most log messages with the assistance of ICL, its unstable outputs and singular focus on the semantics of a single query may lead to the creation of erroneous and inconsistent templates.
For example, when individually parsing two log messages, \fixedwidth{``User admin successfully login in''} and \fixedwidth{``User user1 successfully login in''}, the LLM may erroneously interpret \fixedwidth{``admin''} as a constant part, while considering \fixedwidth{``user1''} as a parameter.
However, when combining these two log messages for analysis, we can ascertain that both \fixedwidth{``admin''} and \fixedwidth{``user1''} are dynamic parameters, indicating the username.

To address this limitation and ensure the consistency of generated templates, we will compare the generated template with the relevant templates during the cache updating operation.
If the newly generated template exhibits high similarity with an existing relevant template, these two log templates may be derived from the same ground-truth template.
We then leverage the new template to refine the relevant template within the \ourtree.
Otherwise, we will insert it as a new template into the \ourtree.
In detail, after getting the newly generated log template $T_a$ from the \llmparser, we first discern whether $T_a$ could potentially belong to the same ground-truth template as any relevant template in the \ourtree.
We calculate the similarities between $T_a$ and all relevant templates \{$T_1, T_2, \cdots, T_n$\} returned by the cache matching operation. 
Given two templates $T_1$ and $T_2$, we split them into a list of tokens, denoted as $L_1$ and $L_2$.
Then, the similarity between $T_1$ and $T_2$ is defined as: $Sim(T_1, T_2) = \frac{2 \times len(LCS(L_1, L_2))}{len(L_1) + len(L_2)}$, where the $LCS$ is the longest common subsequence of two templates.
We choose $T_b$ from all relevant templates, which exhibits the highest similarity with $T_a$.
(1) If the similarity is smaller than the pre-defined threshold (\eg 0.8 in our implementation), it implies that the new template $T_a$ exhibits a low correlation with these relevant templates.
As a result, \nm directly \textit{insert} $T_a$ into the \ourtree.
For example, in Case2 of Fig.~\ref{fig: component-1}, the similarity between $T_a$ and $T_2$ is small, so $T_a$ is inserted into the \ourtree as a new template.
(2) If $Sim(T_a, T_b)$ exceeds the threshold, it indicates that $T_a$ and $T_b$ are highly similar and likely belong to the same ground-truth template.
Such inconsistent templates may be caused by mistakes of LLMs.
Therefore, we \textit{refine} $T_b$ by merging $T_a$ to ensure the consistency.
This is achieved by modifying the path of $T_b$ within the tree of \ourtree, wherein the differing tokens are replaced with the \fixedwidth{``<*>''}.
An example is shown in Case3 of Fig.~\ref{fig: component-1}, the similarity between $T_a$ and $T_3$ is high, so we refine the \fixedwidth{``admin''} node to \fixedwidth{``<*>''}.
This creates a new refined template \fixedwidth{``User <*> successfully log in''}.
In this manner, \nm can adaptively update the \ourtree, utilizing both the answers of the LLM and historical templates within \ourtree, thus enhancing the accuracy of the parsed templates.
Moreover, as the log templates within the \ourtree are considerably fewer than the log messages, and the cache matching operation selectively filters relevant templates, the overhead associated with cache updating is typically minimal.

%% file: Sections/04-evaluation.tex
\section{Experimental Setup}

\subsection{Research Questions}

We evaluate \nm on public large-scale log datasets by answering the following research questions:
\begin{itemize}[leftmargin=*, topsep=0pt]
    \item \textbf{RQ1:} How effective is \nm in parsing log messages?
    \item \textbf{RQ2:} How does each design contribute to \nm?
    \item \textbf{RQ3:} How capable is \nm integrated with different LLMs?
    \item \textbf{RQ4:} How efficient is \nm in processing large-scale log data?
\end{itemize}

\subsection{Datasets and baselines}
\label{sec: datasets}

Our experiments are conducted using Loghub-2.0~\cite{he2020loghub,jiang2023large}, a collection of large-scale datasets for log parsing from LogPAI~\cite{zhu2019tools}.
Loghub-2.0 contains ground-truth templates of 14 log datasets in Loghub~\cite{he2020loghub} from a wide range of systems, including distributed systems, operating systems, and server-side applications.
On average, each dataset in Loghub-2.0 contains 3.6 million log messages, all labeled with ground-truth log templates.
Besides, the total number of log templates is about 3,500.

In accordance with recent benchmark studies~\cite{khan2022guidelines,jiang2023large}, we select four open-source and state-of-the-art log parsers for comparison with our method.
The first two, AEL~\cite{jiang2008abstracting} and Drain~\cite{he2017drain}, are chosen due to their superior performance among all syntax-based log parsers. 
We also choose two latest semantic-based log parsers, UniParser~\cite{liu2022uniparser} and LogPPT~\cite{le2023log}, considering the highest parsing accuracy they have achieved~\cite{jiang2023large}.
To ensure a fair comparison, we use the implementations of all baselines from their replication repositories, choosing the default settings or hyper-parameters.    

\subsection{Metrics}

Following recent studies~\cite{liu2022uniparser,khan2022guidelines,jiang2023large}, we used the following four metrics in our experiments:
\begin{itemize}[leftmargin=*, topsep=0pt]
    \item \textit{Grouping Accuracy (GA):} 
    GA is computed as the ratio of correctly grouped log messages to the total count of log messages.
    A log message is considered to be correctly grouped if and only if its template aligns with the same set of log messages as that of the ground truth.
    \item \textit{F1 score of Grouping Accuracy (FGA):}
    FGA is a template-level metric that focuses on the ratio of correctly grouped templates.
    Specifically, let $N_g$ be the actual correct number of templates in the ground truth, and $N_p$ be the number of templates that are generated by a log parser. If $N_c$ is the number of templates that are correctly parsed by the log parser, then we can compute the Precision of Grouping Accuracy (PGA) as $\frac{N_c}{N_p}$ and the Recall of Grouping Accuracy (RGA) as $\frac{N_c}{N_g}$. The FGA is equal to their harmonic mean, \ie $\frac{2 \times GPA \times RGA}{PGA + RGA}$.
    \item \textit{Parsing Accuracy (PA):}
    PA evaluates the capacity to extract the templates accurately, which is essential to downstream tasks such as anomaly detection~\cite{liu2022uniparser}.
    PA is defined as the proportion of correctly parsed log messages to the total number of log messages.
    A log message is regarded to be correctly parsed if, and only if, all tokens of templates and variables are accurately identified.
    \item \textit{F1 score of Template Accuracy (FTA):}
    Similar to FGA, FTA is a template-level metric that is calculated based on the proportion of correctly identified templates.
    It is computed as the harmonic mean of Precision and Recall of Template Accuracy.
    Differently, a template is regarded as correctly identified if and only if log messages of the parsed template share the same ground-truth template and all tokens of the template are the same as those of the ground-truth template.
\end{itemize}

\subsection{Implementation and Environment}

We conduct our experiments on an Ubuntu 20.04.5 LTS server with 256GB RAM and an NVIDIA GeForce GTX3090 since UniParser and LogPPT require GPU resources to perform log parsing.
The default LLM in \nm is set to ChatGPT (\textit{gpt-3.5-turbo-0613}), primarily due to its popularity in recent research~\cite{peng2023generative,li2023exploring,xu2023prompting,le2023evaluation}.
We call ChatGPT through the official API provided by OpenAI~\cite{openai-api} and set its temperature to 0 so that ChatGPT would generate the same output for the same query to ensure reproducibility. 
Moreover, we also employ different LLMs to explore the generalizability of \nm.
To simulate the practical usage of \nm, we use the sampling algorithm to select candidates from the first 20\% of the log messages in each dataset, and the default number of candidate samples and demonstration examples are set to 32 and 3, respectively.
We also evaluate the performance of \nm with different numbers of sampled candidates and demonstration examples.

We have implemented \nm in Python and integrated it into previous benchmarks~\cite{zhu2019tools,khan2022guidelines,jiang2023large} so that we can fairly compare \nm and all baselines in the same framework.
For all experiments that exhibit randomness, we repeat them five times and report the median results following previous work~\cite{khan2022guidelines,jiang2023large,xu2023prompting} to avoid potential random bias.

\section{Evaluation Results}

\subsection{RQ1: How effective is \nm in parsing log messages?}

\input{Tables/RQ1-effectiveness}

In this RQ, we conduct a comprehensive evaluation of the accuracy and robustness of \nm in comparison to other state-of-the-art baselines on public datasets.

\subsubsection{Accuracy}

The accuracy is the most critical factor in the effectiveness of log parsers.
We employ the default settings of all methods (\eg 32 sampled candidates for both LogPPT and \nm) and apply them to all log datasets.
The four selected metrics are shown in Table.~\ref{tab: RQ1}, and the best results for each metric on each dataset are marked in \textbf{bold} font.
The metrics for AEL on Spark are denoted as ``--'' since it cannot complete the parsing process of the Spark dataset within a reasonable time (\ie 12 hours), following previous works~\cite{khan2022guidelines,jiang2023large}.

According to the evaluation results, it is clear that \nm outperforms all baselines on all average metrics.
In specific, in terms of group-related metrics (\ie GA and FGA), \nm achieves average scores of 92.7\% and 92.4\% on GA and FGA, outperforming Drain by 10.0\% and 66.8\%.
However, the best baseline, Drain, achieves a GA of 84.3\% but only an FGA of 55.4\%.
This is due to the imbalanced frequencies of templates in log datasets, and these log parsers may generate a large number of redundant and erroneous log templates, thereby leading to a markedly low PTA, which subsequently results in a low FTA.
These redundant and erroneous templates can also seriously affect downstream tasks.
Although the inherent instability of generative models causes the grouping-related metrics of UniParser and LogPPT to be inferior to other syntax-based log parsers, the designs of the parsing cache within \nm are able to mitigate this issue, achieving superior grouping metrics.

When considering the metrics related to parsing ability (\ie PA and FTA), LogPPT has achieved the highest PA of 73.6\% and FTA of 47.8\% among all baselines.
However, without the tuning process, \nm has achieved superior parsing metrics, with a PA of 84.2\% and an FTA of 81.0\%, which outperforms LogPPT by 14.4\% and 69.5\%, respectively.
For the most stringent and comprehensive metric, the FTA, \nm surpasses all baselines across all datasets.
Given the strict definitions of correctly parsed and correctly identified, achieving such high metrics signifies that \nm indeed possesses a strong capacity to distinguish between log templates and parameters.

\subsubsection{Robustness}

\begin{figure*}[htbp]
    \vspace{-5pt}
    \centering
    \includegraphics[width=\textwidth]{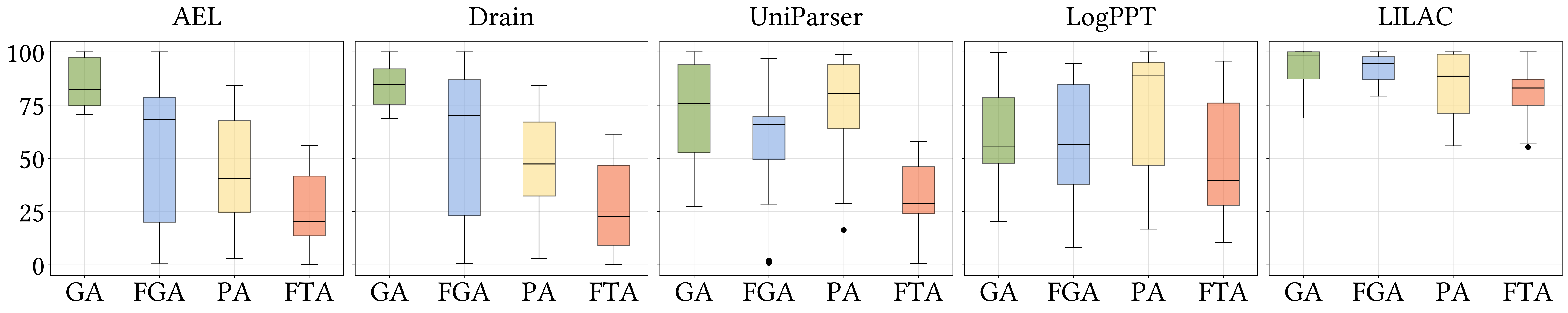}
    \caption{Robustness comparison between baselines and \nm on public datasets (\%)} 
    \label{fig: RQ1-robustness}
    \vspace{-5pt}
\end{figure*}

The robustness of log parsers is also an essential factor in evaluating their effectiveness.
The strong robustness implies that log parsers can maintain a stable performance when dealing with log data of diverse characteristics, indicating a superior generalizability~\cite{zhu2019tools,le2023log,jiang2023large,xu2023prompting}.
To compare the robustness of \nm with all baselines, we draw the box plot illustrating the distribution of each log parser's metrics across all datasets, as depicted in Fig.~\ref{fig: RQ1-robustness}.

It is obvious that \nm not only achieves the highest accuracy but also exhibits the least performance variance, as evidenced by its narrowest distribution range.
This demonstrates that \nm exhibits the strongest robustness when parsing various log data.
Specifically, the standard deviations of \nm for GA, FGA, PA and FTA are 9.3\%, 6.9\%, 15.2\%, and 12.9\%, respectively.
In contrast, these values for LogPPT are 26.0\%, 27.4\%, 27.6\%, and 27.3\%.
The strong robustness of \nm is primarily derived from the vast pre-trained knowledge related to logs of LLMs.
In addition, the ICL paradigm in the \llmparser adapts the LLM to the system-specific characteristics of specific log datasets, thereby enhancing the robustness of parsing diverse log data.

\begin{tcolorbox}[boxsep=1pt,left=2pt,right=2pt,top=3pt,bottom=2pt,width=\linewidth,colback=white!90!black,boxrule=0.2pt,colbacktitle=white!,toptitle=2pt,bottomtitle=1pt,opacitybacktitle=0,breakable]
\textbf{Answer to RQ1:} 
\nm outperforms baseline methods on all metrics, with notable improvements of 66.8\% and 69.5\% for FGA and FTA, respectively, compared to Drain and LogPPT.
Furthermore, \nm exhibits the strongest robustness, reflected in the minimal performance variance when parsing diverse log data from different systems. 
\end{tcolorbox}

\subsection{RQ2: How does each design contribute to \nm?}

In this RQ, we conduct a series of experiments to investigate the contributions of two designed modules within \nm, \ie the \llmparser and the \ourtree.

\subsubsection{ICL-enhanced Parser}

In the \llmparser, we have designed an effective and efficient hierarchical candidate sampling algorithm, along with a kNN-based demonstration selection.
In this section, we aim to investigate the individual contributions of these two designs and explore how different numbers of candidates or demonstrations will affect the performance of \nm. 

\textbf{Contribution of ICL design choices.}
We first assess the individual contributions of the candidate sampling and demonstration selection algorithms.
Specifically, we create the following four variants of \nm and compare them with the original approach.
1) \nm w/o ICL: remove the ICL design in \llmparser,
2) \nm w/ random selection: replace the kNN-based demonstration selection with random selection,
3) \nm w/ random sampling: replace the candidate sampling algorithm with random sampling,
4) \nm w/ LogPPT sampling: replace the candidate sampling algorithm with the adaptive sampling algorithm of LogPPT.

\input{Tables/RQ2-performance}

The evaluation results are depicted in Table.~\ref{tab: RQ3-performance}, in which the following observations can be made.
(1) The absence of the ICL design substantially negatively impacts the performance of \nm across all four metrics.
For instance, the average FTA of \nm experiences a considerable decrease of 27.9\% when the ICL design is removed.
The reason is that even though LLMs possess extensive pre-trained knowledge, their ability to effectively handle a wide range of log data remains limited in the absence of ICL capability.
(2) Upon replacing the hierarchical candidate sampling and kNN-based demonstration selection algorithms with random strategies, there is a respective decrease of 17.7\% and 15.7\% in the average FTA, while the FGA values experience a reduction of 14.6\% and 12.4\%.
This underscores the significance of the quality of candidate samples and demonstrations in influencing the performance of LLMs.
(3) When we replace the original candidate sampling algorithm with that of LogPPT, there is a varying degree of decline across all four metrics, \eg the average FTA is reduced by 14.2\%.
The reason is that the sampling algorithm of LogPPT does not consider the issue of imbalanced template frequencies and the representativeness of the sampled candidates. 
In contrast, our proposed sampling algorithm is capable of sampling diverse and representative candidates, thereby effectively guiding LLMs to accurately parse the entire log dataset.

\textbf{Impact of ICL parameter settings.}
In addition to the above-mentioned individual contributions of the designed algorithms, we also conduct experiments to evaluate the performance of \nm using different numbers of sampled candidates and selected demonstrations.

\begin{figure*}[htbp]
    \centering
    \vspace{-5pt}
    \includegraphics[width=\textwidth]{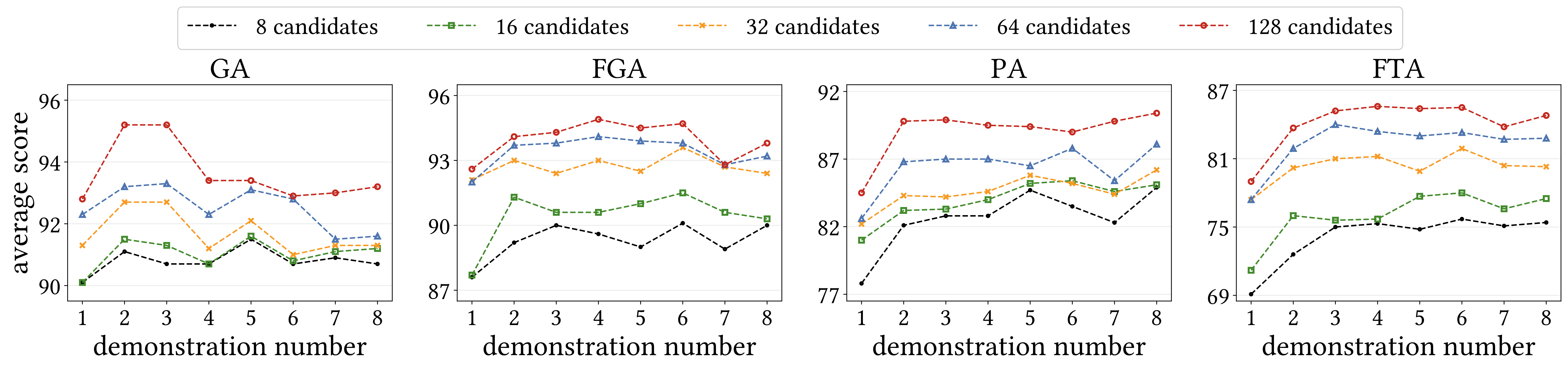}
    \caption{Average accuracy among different numbers of sampled candidates and selected demonstrations. (\%)} 
    \label{fig: RQ2-influence}
    \vspace{-5pt}
\end{figure*}

The results are shown in Fig.~\ref{fig: RQ2-influence}.
It is clear that different numbers of both candidates and demonstrations can affect the performance of \nm.
Specifically,
(1) Even though we only sampled 8 candidates and set the demonstration number greater than 3, the average FTA of \nm is around 75\%, which is much higher than \nm without ICL design in Table.~\ref{tab: RQ3-performance} (\ie 58.4\%).
This suggests that the ICL design can effectively adapt the LLM to parse various log data even when the quantity of labeled data is limited.
(2) As the number of candidates increases, all four metrics of \nm exhibit an improvement.
For instance, when the number of demonstrations is set to 3, the average FTAs for candidate numbers 8, 32, and 128 are approximately 75\%, 81\%, and 85\%, respectively.
This is because the more sampled candidates can provide a broader range of semantic and pattern characteristics in log data, which can be selected to demonstrate LLMs, thereby enabling more precise template generation.
(3) The performance of \nm is influenced by the varying number of demonstrations for each query under each setting of the candidate number.
In particular, the performance of \nm is lowest when only a single demonstration is used since a single demonstration can introduce inductive bias into the parsing process of LLMs.
However, when the number of demonstrations exceeds three and continues to increase, the performance of \nm exhibits fluctuations but tends towards stability.
This is because our kNN-based algorithm is capable of selecting demonstrations that are not only similar to the queried log but also exhibit a high degree of consistency.
In summary, although the performance of \nm is influenced by the varying number of sampled candidates and selected demonstrations, an enhancement in the performance of \nm is observed across all settings when compared to \nm w/o ICL.
Moreover, the most appropriate configuration of 32 candidates and 3 demonstrations is selected as the default setting in our other experiments.

\subsubsection{Parsing Cache}
\label{sec: parsing cache}

\input{Tables/RQ2-ParsingCache}

One of the design objectives of \ourtree is to mitigate the inconsistency in the answers of LLMs.
To validate it, in this section, we evaluate the contribution of the \ourtree to the performance enhancement of \nm.
A direct approach is comparing the performance of the original \nm with that of \nm without the \ourtree.
However, considering the substantial size of these log datasets (averaging 3.6 million log messages per dataset) and the overhead of querying the LLM, it is infeasible to utilize current LLMs for parsing these datasets without the aid of \ourtree, \ie processing line by line.
Instead, we replace the \llmparser with a smaller language model, RoBERTa, which is used by the latest semantic-based log parsers, LogPPT.
Both RoBERTa and LLMs exhibit the common issue of unstable outputs, given that they are both generative language models.
Consequently, comparing the original LogPPT and LogPPT with \ourtree can reflect the effectiveness of \ourtree in mitigating the instability associated with generative language models.

The results are presented in Table.~\ref{tab: RQ2-parsing_cache}.
It is evident that the integration of \ourtree has substantially improved the performance of LogPPT.
First, regarding the grouping-related metrics, the mean GA and FGA of LogPPT with \ourtree have risen by 48.4\% and 55.1\%, respectively, in contrast to the original LogPPT.
This implies that by matching and adaptively updating \ourtree, \nm can ensure the consistency of templates generated by language models, thereby improving the accuracy of grouping.
Second, both the mean PA and FTA have demonstrated respective increases of 6.0\% and 41.0\%.
This suggests that the specifically designed refinements of templates within \ourtree can accurately rectify the incorrect templates produced by language models based on historical templates.

\begin{tcolorbox}[boxsep=1pt,left=2pt,right=2pt,top=3pt,bottom=2pt,width=\linewidth,colback=white!90!black,boxrule=0.2pt,colbacktitle=white!,toptitle=2pt,bottomtitle=1pt,opacitybacktitle=0,breakable]
\textbf{Answer to RQ2:}
Both designs of \llmparser and \ourtree significantly contribute to enhancing \nm's overall performance. 
On the one hand, the proposed ICL strategies provide LLMs the capability of accurately parsing log messages.
On the other hand, \ourtree is effective in mitigating the inconsistency inherent in language models.
\end{tcolorbox}

\subsection{RQ3: How capable is \nm integrated with different LLMs?}

\input{Tables/RQ3-capability-1}

In this RQ, we compare the performance of \nm by employing different LLMs in \llmparser.
Specifically, we select three representative LLMs commonly used in research~\cite{xu2023prompting,gao2023constructing}, namely, ChatGPT, Davinci, and Curie.
Both ChatGPT and Davinci possess a substantial model parameter count of 175B.
ChatGPT, having been fine-tuned for conversational tasks, provides a superior generation speed.
Conversely, Davinci has enhanced capabilities in executing text-generation tasks.
Furthermore, Curie is distinguished by the smallest parameter size, amounting to 13B.

Table.~\ref{tab: RQ3-1} demonstrates the average metrics of \nm with different LLMs, from which we can find consistently high performance across all LLMs. 
In detail, both \nm with ChatGPT and Davinci have achieved exceedingly high average metrics, due to their vast parameter volume and extensive pre-training knowledge.
We have also observed that the majority of these four metrics for Davinci marginally surpass those of ChatGPT, \eg FTA augmented by 0.6\%.
The reason could be that Davinci is more focused on text-generation tasks, which aligns with the log parsing task.
Furthermore, we can observe that the performance of \nm with Curie is the most inferior, \eg the average FTA of Curie is 12.1\% lower than that of ChatGPT.
This is due to the limited model parameters and pre-training knowledge of Curie, signifying a poorer text processing capability, as well as a weaker ICL capacity~\cite{wang2023investigating}.
However, \nm with a comparatively smaller LLM can still achieve an accuracy surpassing all existing log parsers.
These results demonstrate that \nm can be generally applied to different LLMs, maintaining high accuracy.

\begin{tcolorbox}[boxsep=1pt,left=2pt,right=2pt,top=3pt,bottom=2pt,width=\linewidth,colback=white!90!black,boxrule=0.2pt,colbacktitle=white!,toptitle=2pt,bottomtitle=1pt,opacitybacktitle=0,breakable]
\textbf{Answer to RQ3:} 
The performance of \nm can be influenced by the capabilities of LLMs.
Nevertheless, \nm is able to consistently achieve high performance, even when utilizing relatively smaller LLMs.
\end{tcolorbox}

\subsection{RQ4: How efficient is \nm in processing large-scale log data?}

Efficiency is of paramount importance in the practical application of log parsers, given the substantial volume of logs~\cite{zhu2019tools,wang2022spine,le2023log}.
\nm encompasses two primary time costs, \ie the time of the candidate sampling process and the parsing process.
In this RQ, we assess the efficiency of these two procedures within \nm, utilizing the public Loghub-2.0 datasets as described in Sec.~\ref{sec: datasets}.

\input{Tables/RQ4-sampling_efficiency}

\subsubsection{Candidate Sampling Efficiency}

Although \citet{xu2023prompting} have proposed a DPP-based sampling algorithm for log parsing, calculating pair-wise distances between all log messages makes it infeasible for execution on large-scale datasets.
In this section, we perform the sampling algorithm of \nm and LogPPT on all datasets and calculate the average sampling time.
The results are shown in Table.~\ref{tab: RQ4-sampling}, from which we can conclude that the efficiency of the sampling algorithm within \nm significantly surpasses that of the LogPPT.
For instance, when sampling 32 candidates, the algorithm of LogPPT requires more than 1300 seconds, whereas \nm only necessitates 19.3 seconds, achieving a speedup of 67.6.
Moreover, the time cost of the LogPPT sampling algorithm increases linearly with the number of sampled candidates as it employs an iterative approach.
In contrast, the time cost of the \nm sampling algorithm remains stable regardless of the number of candidates sampled.
The reason is that the time overhead of \nm's sampling algorithm is almost on hierarchical clustering, which is efficient for handling extensive log data.

\subsubsection{Parsing Efficiency}

\begin{figure*}[htbp]
    \vspace{-6pt}
    \centering
    \includegraphics[width=0.7\textwidth]{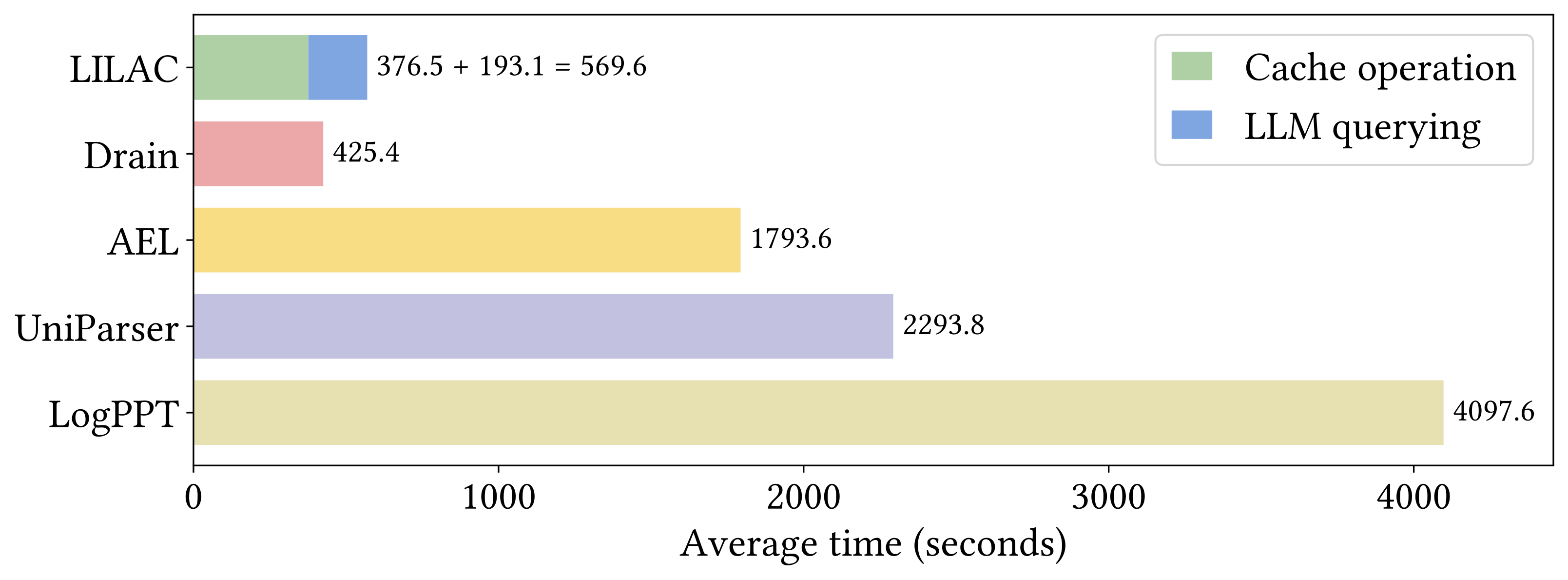}
    \vspace{-5pt}
    \caption{Efficiency of baselines and \nm on large-scale datasets} 
    \label{fig: RQ4-efficiency}
    \vspace{-6pt} 
\end{figure*}

In this section, our primary focus is on assessing the efficiency of the parsing process within \nm.
More specifically, we have recorded the execution times for all baselines and \nm with ChatGPT on all log datasets.
Furthermore, we have separately recorded the cache operation time of the \ourtree within \nm and the time expended on querying the \llmparser.
Specifically, the cache operation time encompasses the time cost for cache matching and cache updating operations, whereas the query time represents the total duration from initiating a query to the \llmparser to receiving the generated templates.
We calculate the average parsing time across all log datasets and plot a bar chart.
The detailed parsing times are available in our replication package~\cite{repo}.
According to the results in Fig.~\ref{fig: RQ4-efficiency}, we can see that \nm exhibits efficiency comparable to that of Drain, the most efficient syntax-based log parser currently available.
In detail, \nm requires 569.6 seconds to process an average of 3.6 million log messages, while this time of Drain is 425.4.
Conversely, other semantic-based log parsers, including UniParser and LogPPT, despite the utilization of GPU acceleration, only achieve low efficiencies, trailing \nm by 4.03 and 7.19 times, respectively.

Besides, across all datasets, the average processing time for \ourtree in \nm is 376.5s, accounting for approximately 66.1\% of the total time.
This is less than the average processing time of Drain, suggesting that the cache matching and updating operations of the \ourtree are highly efficient.
Correspondingly, the time consumed by querying the \llmparser averages at 193.1s, representing 33.9\% of the total time.
We further conduct a statistical analysis on the number of queries to the LLM.
The mean value of query numbers is 279.7, while the average number of ground-truth templates is 249.
The observed difference is caused by the incorrect templates generated by the LLM, which results in failed matching of the \ourtree and consequently leads to unnecessary queries.
However, \nm can effectively keep this number minimal, thereby ensuring efficiency.
When compared with the existing LLM-based log parsing approaches~\cite{xu2023prompting,le2023evaluation,liu2023logprompt}, which necessitate over 3.6 million queries, \nm markedly diminishes the number of queries to LLMs.
This makes the application of LLMs for log parsing practically feasible.

\begin{tcolorbox}[boxsep=1pt,left=2pt,right=2pt,top=3pt,bottom=2pt,width=\linewidth,colback=white!90!black,boxrule=0.2pt,colbacktitle=white!,toptitle=2pt,bottomtitle=1pt,opacitybacktitle=0,breakable]
\textbf{Answer to RQ4:} 
\nm substantially reduces the number of queries to the LLM by matching the \ourtree to prevent duplicate queries.
Hence, the efficiency of \nm surpasses semantic-based methods by 4.03 to 7.19 times and is comparable to the fastest syntax-based methods.
\end{tcolorbox}

%% file: Tables/RQ1-effectiveness.tex
\begin{table*}[htbp]
\captionsetup{justification=centering}
\centering
\vspace{-5pt}
\caption{Accuracy comparison between baselines and \nm  on public datasets (\%)}
\label{tab: RQ1}
 \resizebox{\textwidth}{!}{
\begin{tabular}{c|cccc|cccc|cccc|cccc|cccc}
\toprule
\  & \multicolumn{4}{c|}{AEL} & \multicolumn{4}{c|}{Drain} & \multicolumn{4}{c|}{UniParser} & \multicolumn{4}{c|}{LogPPT} & \multicolumn{4}{c}{\nm} \\
\  & GA & FGA & PA & FTA & GA & FGA & PA & FTA & GA & FGA & PA & FTA & GA & FGA & PA & FTA & GA & FGA & PA & FTA \\
\hline
Hadoop & 82.3 & 11.7 & 53.5 & 5.8 & \textbf{92.1} & {78.5} & 54.1 & 38.4 & 69.1 & 62.8 & \textbf{88.9} & {47.6} & 48.3 & 52.6 & 66.6 & 43.4 & {87.2} & \textbf{96.2} & {83.2} & \textbf{77.9} \\ 
HDFS & 99.9 & 76.4 & 62.1 & 56.2 & 99.9 & 93.5 & 62.1 & {60.9} & \textbf{100} & \textbf{96.8} & {94.8} & 58.1 & 72.1 & 39.1 & 94.3 & 31.2 & \textbf{100} & \textbf{96.8} & \textbf{99.9} & \textbf{94.6} \\ 
OpenStack & 74.3 & 68.2 & 2.9 & 16.5 & 75.2 & 0.7 & 2.9 & 0.2 & \textbf{100} & {96.9} & {51.6} & 28.9 & 53.4 & 87.4 & 40.6 & {73.8} & \textbf{100} & \textbf{100} & \textbf{100} & \textbf{97.9} \\ 
Spark & --- & --- & --- & --- & {88.8} & {86.1} & 39.4 & {41.2} & 85.4 & 2.0 & 79.5 & 1.2 & 47.6 & 37.4 & {95.2} & 29.9 & \textbf{100} & \textbf{90.1} & \textbf{97.3} & \textbf{75.9} \\ 
Zookeeper & {99.6} & 78.8 & 84.2 & 46.5 & 99.4 & 90.4 & 84.3 & 61.4 & 98.8 & 66.1 & \textbf{98.8} & 51.0 & 96.7 & {91.8} & {84.5} & {80.9} & \textbf{100} & \textbf{96.7} & 68.7 & \textbf{86.8} \\ 
BGL & 91.5 & 58.7 & 40.6 & 16.5 & \textbf{91.9} & {62.4} & 40.7 & 19.3 & {91.8} & {62.4} & {94.9} & 21.9 & 24.5 & 25.3 & 93.8 & {26.1} & 89.4 & \textbf{85.9} & \textbf{95.8} & \textbf{74.6} \\ 
HPC & 74.8 & 20.1 & 74.1 & 13.6 & {79.3} & 30.9 & 72.1 & 15.2 & 77.7 & 66.0 & {94.1} & 35.1 & 78.2 & {78.0} & \textbf{99.7} & {76.8} & \textbf{86.9} & \textbf{90.7} & 70.5 & \textbf{80.0} \\ 
Thunderbird & 78.6 & 11.6 & 16.3 & 3.5 & \textbf{83.1} & 23.7 & 21.6 & 7.1 & 57.9 & {68.2} & \textbf{65.4} & {29.0} & 56.4 & 21.6 & 40.1 & 11.7 & {80.6} & \textbf{79.3} & {55.9} & \textbf{57.2} \\ 
Linux & {91.6} & {80.6} & 8.2 & 21.7 & 68.6 & 77.8 & 11.1 & 25.9 & 28.5 & 45.1 & 16.4 & 23.2 & 20.5 & 71.2 & {16.8} & {42.8} & \textbf{97.1} & \textbf{93.1} & \textbf{76.5} & \textbf{74.0} \\ 
Mac & {79.7} & {79.3} & 24.5 & 20.5 & 76.1 & 22.9 & 35.7 & 6.9 & 73.7 & 69.9 & \textbf{68.8} & {28.3} & 54.4 & 49.3 & 39.0 & 27.4 & \textbf{87.6} & \textbf{82.5} & {63.8} & \textbf{55.3} \\ 
Apache & \textbf{100} & \textbf{100} & 72.7 & {51.7} & \textbf{100} & \textbf{100} & 72.7 & {51.7} & 94.8 & 68.7 & 94.2 & 26.9 & 78.6 & 60.5 & {94.8} & 36.8 & \textbf{100} & \textbf{100} & \textbf{99.6} & \textbf{86.2} \\ 
OpenSSH & {70.5} & 68.9 & 36.4 & 33.3 & \textbf{70.7} & \textbf{87.2} & 58.6 & {48.7} & 27.5 & 0.9 & 28.9 & 0.5 & 27.7 & 8.1 & {65.4} & 10.5 & 69.0 & {83.8} & \textbf{94.1} & \textbf{86.5} \\ 
HealthApp & 72.5 & 0.8 & 31.1 & 0.3 & 86.2 & 1.0 & 31.2 & 0.4 & 46.1 & 74.5 & {81.7} & 46.2 & {99.8} & {94.7} & \textbf{99.7} & {82.2} & \textbf{100} & \textbf{98.1} & 72.9 & \textbf{87.3} \\ 
Proxifier & 97.4 & 66.7 & 67.7 & 41.7 & 69.2 & 20.6 & 68.8 & 17.6 & 50.9 & 28.6 & 63.4 & 45.7 & {98.9} & {87.0} & \textbf{100} & {95.7} & \textbf{100} & \textbf{100} & \textbf{100} & \textbf{100} \\ 
\hline
Average & {85.6} & 55.5 & 44.2 & 25.2 & 84.3 & 55.4 & 46.8 & 28.2 & 71.6 & {57.8} & 73.0 & 31.7 & 61.2 & 57.4 & {73.6} & {47.8} & \textbf{92.7} & \textbf{92.4} & \textbf{84.2} & \textbf{81.0} \\ 
\bottomrule
\end{tabular}
\vspace{-5pt}
}
\end{table*}

%% file: Tables/RQ2-performance.tex
\begin{table*}[htbp]
\vspace{-5pt}
\captionsetup{justification=centering}
\centering
\caption{Average accuracy comparison among \nm with different strategies (\%)}
\label{tab: RQ3-performance}
 \resizebox{0.8\textwidth}{!}{
\begin{tabular}{lcccc}
\toprule
 & GA & FGA & PA & FTA \\
\hline
\nm & 92.7 & 92.4 & 84.2 & 81.0 \\
w/o ICL & 83.5 ($\downarrow 9.9\%$) & 76.5 ($\downarrow 17.2\%$) & 62.6 ($\downarrow 25.6\%$)& 58.4 ($\downarrow 27.9\%$)\\
w/ random selection &  84.2 ($\downarrow 9.2\%$) & 80.6 ($\downarrow 14.6\%$) & 74.4 ($\downarrow 11.6\%$) & 66.7 ($\downarrow 17.7\%$)\\
w/ random sampling & 87.6 ($\downarrow 5.5\%$) & 80.9 ($\downarrow 12.4\%$) & 77.5 ($\downarrow 8.0\%$)&  68.3 ($\downarrow 15.7\%$)\\
w/ LogPPT sampling & 91.3 ($\downarrow 1.5\%$) & 84.8 ($\downarrow 8.2\%$) & 79.7 ($\downarrow 5.3\%$)& 74.9 ($\downarrow 7.5\%$)\\
\bottomrule
\end{tabular}
}
\vspace{-5pt}
\end{table*}

%% file: Tables/RQ2-ParsingCache.tex
\begin{table*}[htbp]
\vspace{-8pt}
\captionsetup{justification=centering}
\centering
\caption{Average accuracy comparison between LogPPT and LogPPT with \ourtree (\%)}
\label{tab: RQ2-parsing_cache}
 \resizebox{0.75\textwidth}{!}{
\begin{tabular}{lcccc}
    \toprule
     & GA & FGA & PA & FTA \\
     \hline
    LogPPT (original) & 61.2 & 57.4 & 73.6 & 47.8 \\
    w/ \ourtree & 90.8 ($\uparrow 48.4\%$) & 89.0 ($\uparrow 55.1\%$) & 78.0 ($\uparrow 6.0\%$) & 67.4 ($\uparrow 41.0\%$) \\
    \bottomrule
\end{tabular}
}
\vspace{-8pt}
\end{table*}

%% file: Tables/RQ3-capability-1.tex
\begin{table*}[]
\vspace{-4pt}
\captionsetup{justification=centering}
\centering
\caption{Average accuracy comparison among \nm with different LLMs (\%)}
\label{tab: RQ3-1}
 \resizebox{0.7\textwidth}{!}{
\begin{tabular}{lcccc}
\toprule
 & GA & FGA & PA & FTA \\
\hline
ChatGPT & 92.7 & 92.4 & 84.2 & 81.0 \\
Davinci & 91.9 ($\downarrow 0.9\%$) & 92.9 ($\uparrow 0.5\%$) & 87.1 ($\uparrow 3.4\%$)& 81.5 ($\uparrow 0.6\%$)\\
Curie & 90.1 ($\downarrow 2.8\%$) & 87.6 ($\downarrow 5.2\%$) & 77.8 ($\downarrow 7.6\%$) & 71.2 ($\downarrow 12.1\%$) \\
\bottomrule
\end{tabular}
}
\vspace{-4pt}
\end{table*}


%% file: Tables/RQ4-sampling_efficiency.tex
\begin{table*}[htbp]
\vspace{-4pt}
\captionsetup{justification=centering}
\centering
\caption{Average sampling time of \nm and LogPPT algorithms on large-scale datasets (seconds)}
\label{tab: RQ4-sampling}
 \resizebox{0.85\textwidth}{!}{
\begin{tabular}{lccccc}
\toprule
 & 8 candidates & 16 candidates & 32 candidates & 64 candidates & 128 candidates\\
 \hline
LogPPT & 284.1 & 629.3 & 1303.9 & 2779.6 & 5396.9 \\
\nm & 19.2 & 19.3 & 19.3 & 19.3 & 19.4 \\
 \hline
Speed up ($\uparrow$) & 14.8 $\times$ & 32.6 $\times$ & 67.6 $\times$ & 144.0 $\times$ & 278.2 $\times$ \\
\bottomrule
\end{tabular}
}
\vspace{-4pt}
\end{table*}

%% file: Sections/05-discussion.tex
\section{Discussion}

\subsection{Practicality of \nm}

\nm is designed to leverage the power of LLMs for log parsing in production systems.
To reduce the cost of querying LLMs and alleviate the inherent instability of query results, \nm adopts the adaptive \ourtree.
In addition, for each query, since the number of tokens in a single log message or template is generally small (\eg tens to hundreds of tokens), \nm would not incur a substantial cost of querying LLMs.
Additionally, \nm introduces effective candidate sampling and demonstration selection algorithms to facilitate the ICL capability of LLMs in log parsing.

It is worth noting that \nm is compatible with traditional language models, such as RoBERTa.
According to the experimental results in Sec.~\ref{sec: parsing cache}, \nm integrated with traditional language models can also achieve higher performance than state-of-the-art log parsing methods.
When using traditional language models, users can utilize the proposed candidate sampling algorithm to obtain high-quality data for model training or tuning.
We believe the above features make \nm a practical framework that can be deployed in real-world systems.

\vspace{-5pt}
\subsection{Threats to Validity}

\noindent
\textbf{Data Leakage.}
Since LLMs are trained on huge volumes of data, one potential threat is the data leakage problem.
Particularly, the adopted LLM in \nm may have been trained on open-source log datasets, leading to the memorization of ground-truth templates as opposed to performing inference.
However, according to our experiments, the performance of \nm without ICL is significantly inferior to \nm with ICL, implying a low probability of direct memorization.
Furthermore, \nm employs the \textit{gpt-turbo-3.5-0613} model for most of the experiments.
It is noteworthy that updates for this model were discontinued before the ground-truth templates in Loghub-2.0 were publicly available.
Therefore, the probability of data leakage within our experiment is negligible.

\noindent
\textbf{Privacy Issue.}
From the perspective of enterprises, log messages are sensitive data, as they often encompass a substantial amount of customer and service information.
Employing external LLMs to process internal log data may pose risks to privacy and security problems.
Actually, \nm is a general framework that can support a variety of language models.
Users can integrate their own language models into \nm, thereby avoiding privacy issues.

\noindent
\textbf{Manual Labeling Effort.}
To utilize the ICL capability of LLMs, manual annotation is required to provide the ground-truth templates of the sampled log messages.
To alleviate the labeling effort associated with ICL, we propose an efficient candidate sampling algorithm designed to sample a compact set of diverse and representative log messages.
Our experimental results proved that even with a small number of labeled log messages (\eg 32), \nm can yield a significantly improved performance.

%% file: Sections/06-related.tex
\vspace{-2pt}
\section{Related Work}

Log parsing has emerged as an active research topic in recent years~\cite{zhu2019tools,he2016evaluation,khan2022guidelines,jiang2023large}.
Existing log parsers can be divided into two groups: syntax-based and semantic-based log parsers.
In specific, syntax-based log parsers can be further subdivided into three categories.
\textit{(1) Frequency-based parsers:}
These log parsers~\cite{vaarandi2003data,nagappan2010abstracting,vaarandi2015logcluster,dai2020logram} utilize frequent patterns of token position or n-gram information to distinguish the templates and parameters in log messages.
\textit{(2) Similarity-based parsers:}
These log parsers~\cite{shima2016LenMa,hamooni2016logmine,tang2011logsig} compute similarities between log messages to cluster them into different groups and then extract the constant parts of log messages.
\textit{(3) Heuristics-based parsers:}
These log parsers~\cite{jiang2008abstracting,he2017drain,makanju2009clustering,messaoudi2018search,mizutani2013SHISO,du2016spell,wang2022spine} employ various heuristic algorithms or data structures to identify the log templates based on designed characteristics.
Semantic-based log parsers can achieve higher parsing accuracy by mining semantics from log messages, which is crucial in some downstream tasks~\cite{huo2023semparser,li2023did}.
These methods typically necessitate labeled log data for model training or tuning.
To be precise, a subset of these log parsers~\cite{liu2022uniparser,huo2023semparser,li2023did} formulate log parsing as a token classification problem, employing bidirectional long short-term memory for training.
In addition, LogPPT~\cite{le2023log} tunes a pre-trained language model (\eg RoBERTa) to perform log parsing.

However, recent benchmark studies~\cite{khan2022guidelines,jiang2023large} have identified that the performance of these log parsers is found to be inadequate when dealing with large-scale, complex log data.
This observation motivates our work, which aims to utilize the capabilities of LLMs for more accurate log parsing.
Recently, several studies have been conducted to explore the utilization of LLMs for log analysis, specifically log parsing.
The study by \citet{le2023evaluation} is the pioneer in investigating the performance of LLMs in log parsing, which demonstrates the potential of LLMs in accomplishing log parsing.
\citet{xu2023prompting} propose LogDiv, a method that leverages the ICL capability of LLMs to achieve more accurate log parsing.
To adpot the ICL paradigm, LogDiv transforms all log messages into embeddings and computes pair-wise distances to sample log messages for demonstration, which is infeasible when dealing with an enormous volume of log messages.
Besides, these existing methods solely employ LLMs to sequentially parse each log in a single query.
Hence, they do not address the challenges of efficiency and consistency inherent in utilizing LLMs to log parsing.
This makes them impractical for utilization in real-world scenarios.
In contrast, our proposed method, \nm, addresses these issues by combining the adaptive \ourtree with the \llmparser, enabling accurate and efficient LLM-based log parsing.

%% file: Sections/07-conclusion.tex
\section{Conclusion}

In this paper, we present \nm, a practical log parsing framework using LLMs with adaptive parsing cache.
To utilize the ICL capability to adapt LLMs to parse various log data, \nm adopts effective and efficient candidate sampling and demonstration selection algorithms to select high-quality demonstrations.
Besides, \nm employs the adaptive \ourtree to store log templates, and specifically tailored cache matching and adaptive updating operations help mitigate the inherent inconsistency and inefficiency of LLMs. 
Extensive experiments on large-scale log datasets demonstrate that \nm outperforms all state-of-the-art baselines with high efficiency.
We believe that \nm would benefit both practitioners and researchers in the field of log analysis.

\section*{Acknowledgment}

The work described in this paper was supported by the Research Grants Council of the Hong Kong Special Administrative Region, China (No. CUHK 14206921 of the General Research Fund). We extend our sincere gratitude to the anonymous reviewers for their constructive feedback.